\documentclass{article}
\usepackage{arxiv}

\usepackage[utf8]{inputenc} 
\usepackage[T1]{fontenc}    
\usepackage{hyperref}       
\usepackage{url}            
\usepackage{booktabs}       
\usepackage{amsfonts}       
\usepackage{nicefrac}       
\usepackage{microtype}      
\usepackage{graphicx}
\usepackage{natbib}
\usepackage{doi}
\usepackage{lineno}
\usepackage{nicematrix}
\usepackage{array, longtable, booktabs, makecell}
\usepackage{wasysym}
\usepackage{multicol}
\usepackage{subfig}
\usepackage{siunitx}

\title{Optics design and correction challenges for the high energy booster of FCC-ee}

\date{\today}
\renewcommand{\shorttitle}{\textit{arXiv} ICFA 2023 Newsletter pre-print}

\author{ \href{https://orcid.org/0000-0002-6284-0086}{\includegraphics[scale=0.06]{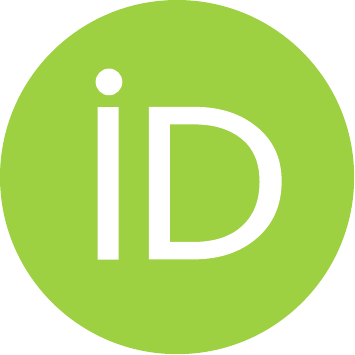}\hspace{1mm}Antoine Chance}\\
	Institut de Recherche sur les lois \\Fondamentales de l'Univers (IRFU)\\
	Commissariat à l'énergie atomique\\ et aux énergies alternatives (CEA)\\
	91191, Gif-sur-Yvette, France \\
	\texttt{antoine.chance@cea.fr} \\
	\And
	\href{https://orcid.org/0000-0002-6808-2810}{\includegraphics[scale=0.06]{orcid.pdf}\hspace{1mm}Barbara Dalena} \\
	Institut de Recherche sur les lois \\Fondamentales de l'Univers (IRFU)\\
	Commissariat à l'énergie atomique\\ et aux énergies alternatives (CEA)\\
	91191, Gif-sur-Yvette, France \\
	\texttt{barbara.dalena@cea.fr} \\
	\And
	Tatiana Da Silva \\
	Institut de Recherche sur les lois \\Fondamentales de l'Univers (IRFU)\\
	Commissariat à l'énergie atomique\\ et aux énergies alternatives (CEA)\\
	91191, Gif-sur-Yvette, France \\
	\texttt{tatiana.dasilva@icloud.com} \\
	\And
	Ahmad Mashal \\
	School of Particles and Accelerators\\
	Institute for Research in \\ Fundamental Sciences (IPM)\\
	P.O. Box 19395-5531, Tehran, Iran\\
	\texttt{ahmad.mashal@cern.ch} \\
	\And
	\href{https://orcid.org/0000-0001-7129-7348}{\includegraphics[scale=0.06]{orcid.pdf}\hspace{1mm}Mauro Migliorati} \\
	University of Rome La Sapienza\\
	NFN Roma1\\
	Piazzale Aldo Moro, 5, 00185 Roma RM - Italy \\
	\texttt{mauro.migliorati@uniroma1.it} \\
	\And
	\href{https://orcid.org/0000-0001-9634-4526}{\includegraphics[scale=0.06]{orcid.pdf}\hspace{1mm}Adnan Ghribi}\thanks{also at GANIL/CNRS/IN2P3 - adnan.ghribi@ganil.fr} \\
	Institut de Recherche sur les lois \\Fondamentales de l'Univers (IRFU)\\
	Commissariat à l'énergie atomique\\ et aux énergies alternatives (CEA)\\
	91191, Gif-sur-Yvette, France \\
	\texttt{adnan.ghribi@cea.fr} \\
	\And
	Ali Rajabi \\
	Deutsches Elektronen-Synchrotron (DESY)\\ Notkestr,
	85, 22607 Hamburg, Germany\\
	\texttt{ali.rajabi@desy.de} \\
	\And
	\href{https://orcid.org/0000-0001-9787-8917}{\includegraphics[scale=0.06]{orcid.pdf}\hspace{1mm}Frank Zimmermann} \\
	European Organization for Nuclear Research (CERN)\\
	1211 Geneva 23 - Switzerland\\
	\texttt{frank.zimmermann@cern.fr} \\
}

\hypersetup{
pdftitle={pre-print for the ICFA newsletter 2023},
pdfsubject={FCC-ee HEB},
pdfauthor={Adnan Ghribi},
pdfkeywords={Accelerator modelling and simulations, Beam dynamics, Beam Optics.},
}

\begin{document}
\maketitle
\shorttitle
\begin{abstract}
One of the major upcoming challenges in particle physics is achieving precise measurements of the Z, W, and H bosons, as well as the top quark. To meet these targets, the next e\textsuperscript{+}e\textsuperscript{-} collider complex, FCC-ee, will need to achieve unprecedented luminosities. The FCC-IS European Study is investigating the feasibility of these challenges, with a cornerstone of the study being the design and optimization of the high-energy booster (HEB). This paper provides an update on the status of the HEB of FCC-ee in light of recent developments in the injector and collider survey, as well as an overview of ongoing work on longitudinal stability and design robustness in relation to field, alignment, and diagnostics errors. Constraints and effects related to the design frequency of the accelerating cavities, as well as collective effects, are also highlighted. Lastly, the paper presents an investigation into an alternative arcs cell design.
\end{abstract}

\section{Introduction}
FCC-ee is a double-ring lepton collider and the first operational stage of the integrated long-term FCC project. It is expected to serve as an electron-positron Higgs and electroweak factory, achieving the highest luminosities within 15 years. There are four operational modes defined for FCC-ee, which are referred to as $Z$, $W$, $H$, and $t\bar{t}$. The beam properties, including energy, current, and emittance, vary for each mode. However, the short beam lifetime and the requirements for top-up injection are common features among them.

The High Energy Booster (HEB) is the final part of the injector complex, where particles with an energy of 20 GeV are injected into the HEB ring. The main criteria for the HEB lattice and its ramping cycle design are accelerating the particles up to the collision energy, adjusting beam properties for efficient top-up injection, and meeting filling time considerations.

Previous studies~\cite{benedikt2018fcc} have shown that a single design could not satisfy the mentioned requirements for all energies. Therefore, two distinguished lattices have been designed, one for $Z$ and $W$ modes and another for $H$ and $t\bar{t}$ modes. Preliminary results confirmed the lattices' performance in the ideal case. However, in the realistic case, the HEB performance is affected by inevitable errors such as magnetic field imperfection and misalignment, instabilities raised from collective effects, and the behavior of particles during the ramping process. Hence, it is necessary to ensure the possibility of correcting destructive effects before finalizing the design.

The effects of magnetic multipole errors on lattice momentum acceptance and a first analytical estimation of emittance growth in the presence of intra-beam scattering have been previously reported \cite{dalena2022status}. This paper is a status report on the aforementioned developments for the HEB design. The second section reports on the study of longitudinal phase space stability, while the third section details closed orbit correction strategies in the presence of misalignment errors. The fourth section shows the beginning of a joint effort for precise collective effects studies. Section six provides an update on the change in RF cavities frequencies with respect to the Conceptual Design Report and the corresponding RF budget. Finally, an alternative optics design under study is presented in the last section.

\section{Optics stability} 
\begin{figure}[h]
    \centering
    \includegraphics[width=1\columnwidth]{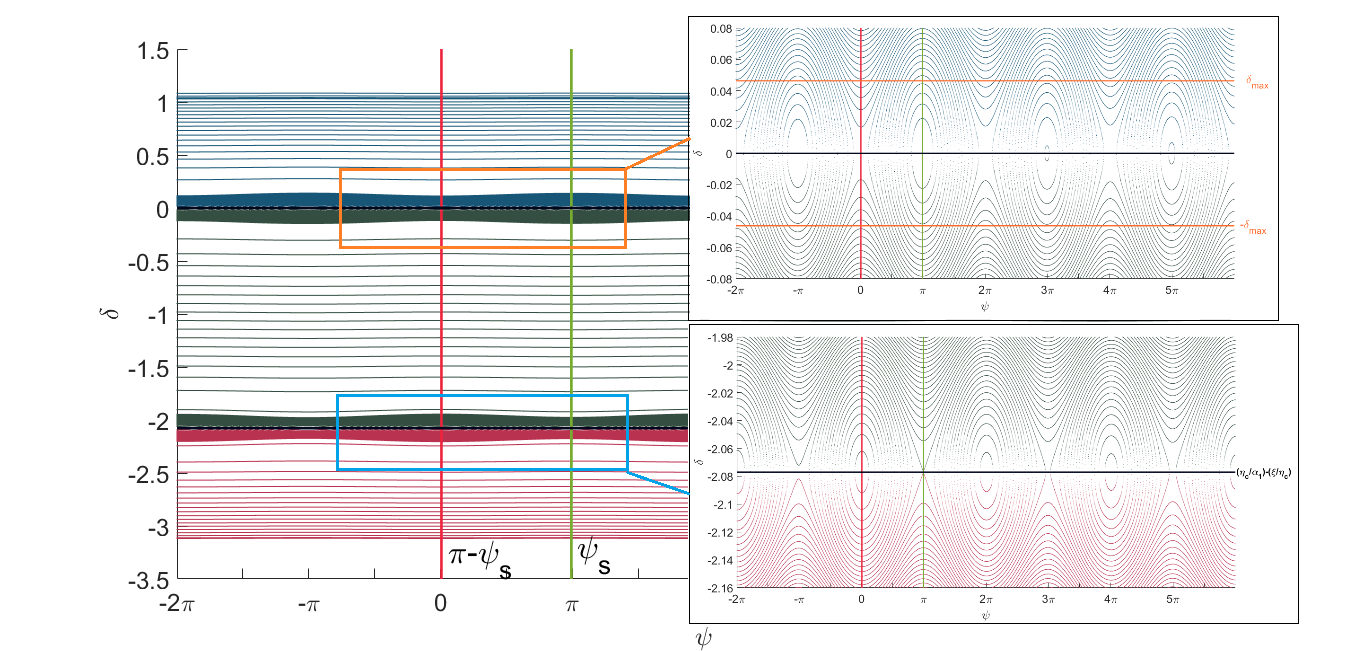}
    \caption{The location of primary and secondary RF buckets in longitudinal phase space.}
    \label{fig:long_space_bucket}
\end{figure}

One of the ongoing tasks is to achieve sufficiently large Dynamic Aperture (DA) and Momentum Aperture (MA) for a \ang{90} phase advance lattice, designed for modes $H$, $tt$ and $t\bar{t}$. By using a non-interleave sextupoles arrangement with 32 pairs of focusing sextupoles and 33 pairs of defocusing sextupoles in each arc, a considerably large dynamic aperture (in absence of the radiation and RF cavities) is obtained. Increasing the number of sextupole families to four and optimizing resonance driving terms enhances the DA, especially for off-momentum particles. The size of both the DA and MA is reduced when a 6D tracking procedure includes radiation effects and energy compensation at RF cavities has been performed. As a first step in resolving this issue, we investigate particles' stability in the longitudinal phase space.
A kilometer-scale bending radius of the dipole magnets, as well as the low value of dispersion function along them, makes the lattice's momentum compaction factor to become very small. In this case, even a slight perturbation in momentum compaction could lead to instability in the longitudinal phase space. A variation in path length with momentum at higher order could be used to compute perturbation terms \cite{wiedemann} as Eq.~\eqref{eq:deltal} shows:
\begin{equation}
\frac{\Delta L}{L_0} =\alpha_c \delta+\alpha_1\delta^2+\xi+\mathcal{O}(3) \label{eq:deltal}
\end{equation}
where $L_0$ is the nominal path length, $\alpha_c$ the momentum compaction factor, $\alpha_1$ the second order momentum compaction and $\xi$ the momentum independent term (see \cite{wiedemann} for more details).
The extended form of the momentum compaction in the Hamiltonian reveals the existence of the secondary RF bucket in the longitudinal phase space. For certain condition which depends on perturbation terms $\alpha_1$ and $\xi$, these buckets could get close and interfere with each other. The buckets’ interference disturbs the longitudinal phase stability condition and reduces the momentum acceptance of the lattice. Moreover, the width of the RF buckets determines by the desired momentum acceptance and its required RF voltage. Hence, the stability criterion for the perturbation term $\alpha_1$ as a function of desired momentum acceptance could be defined \cite{wiedemann} as in Eq.~\eqref{eq:stability_condition}:
\begin{align}
    \alpha_1 &\leq \frac{|\eta_c | (1-\Gamma)^{3⁄4}}{\sqrt{3} (\Delta p/p_0 )_\text{desired}}
    &\text{Where } \Gamma&=\frac{4 \xi \alpha_1}{\eta_c^2}
    \label{eq:stability_condition}
\end{align}
Tracking the on-axis particle within the momentum acceptance range of the ideal lattice gives us $\alpha_c = \num{7.33e-6}$, $\alpha_1 = \num{3.52e-6}$, and $\xi = \num{3.07e-11}$ ; so, the right hand side of Eq.~\eqref{eq:stability_condition} for desired $\pm 5 \%$ momentum acceptance is equal to \num{8.46e-5} and the lattice meets this stability condition. The location of the primary and secondary RF buckets in the longitudinal phase space is shown in figure~\ref{fig:long_space_bucket}. It should be noted that the secondary RF buckets are situated at a considerable distance from the main buckets, and their influence on the particle dynamics can be disregarded.

\begin{figure}[h]
    \centering
    \includegraphics[width=0.8\columnwidth]{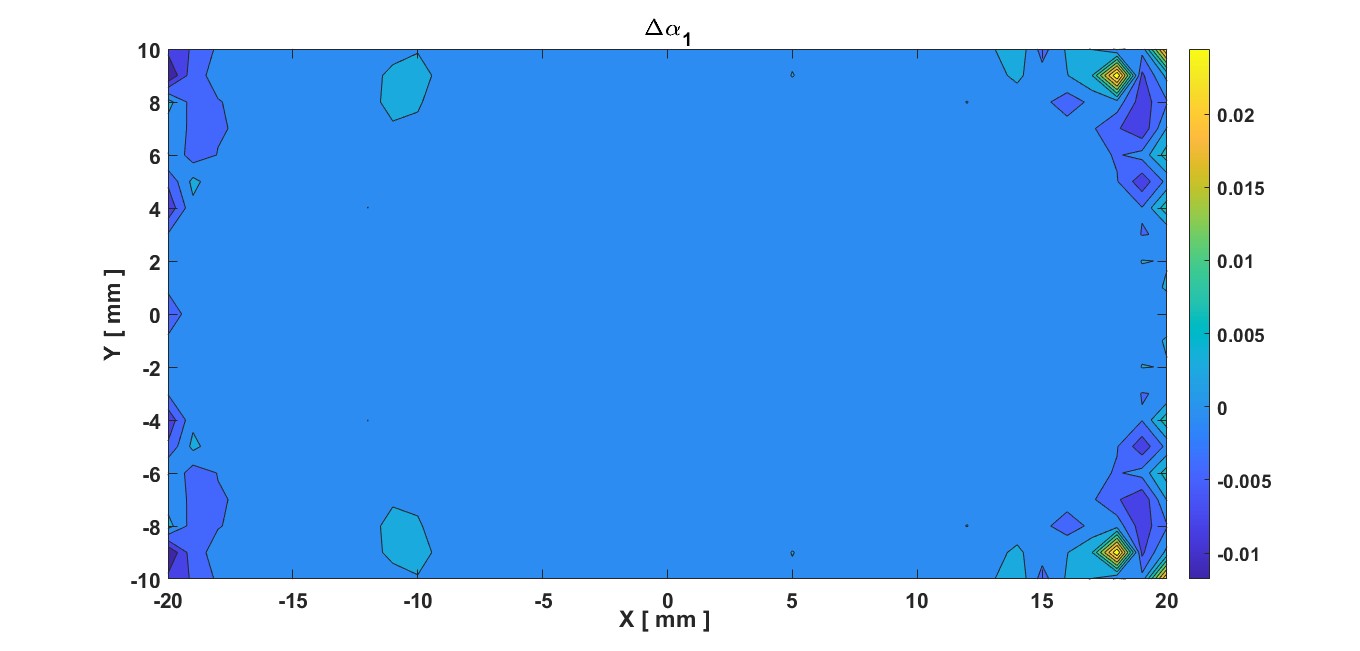}
    \caption{The variation of $\alpha_1$ for the different amplitudes.}
    \label{fig:dyn_stability}
\end{figure}

The initial amplitude of the particles could alter the value of $\alpha_1$, accordingly, the stability criteria for variation of $\alpha_1$ has been defined in \cite{wiedemann} as : 
$\Delta \alpha_1<\frac{\eta_c^2}{4\xi}=0.226$.
To ensure the dynamic stability of the beam, the variation of $\alpha_1$ within the range of $(x, y) \in [-20, 20] \times [-10,10]$ [mm$^2$] has been computed and its result is shown in figure~\ref{fig:dyn_stability}. It could be concluded that the variation of $\alpha_1$ for the defined range is far below the stability threshold.

\section{Orbit correction}
The objective of the orbit correction study is twofold: firstly, to establish the permissible tolerances for the misalignment of the booster's components, and secondly, to assess the necessary strength of the correctors to rectify the closed orbit perturbation resulting from machine errors. Various types of errors have been taken into account, including random dipole field and roll errors, quadrupole alignment errors, BPM alignment and reading errors, and sextupole alignment errors.

The correction strategy described in Ref.~\cite{ipac_orbcor} was employed to evaluate the tolerances for misalignment and field errors on the elements. To do so, we conducted several tests by gradually introducing different error types, each on 100 seeds. The initial test configuration consisted of the combination of quadrupole offsets (MQ), dipole relative field error (MB), and main dipole roll error. Statistical analysis of this initial configuration revealed that all seeds converged until an MQ offset of \SI{150}{\micro\metre} was reached. Notably, all errors applied to the elements were randomly Gaussian distributed within $\pm$ 3 RMS (Root Mean Square).

\begin{figure}
    \centering
    \includegraphics[width=\columnwidth]{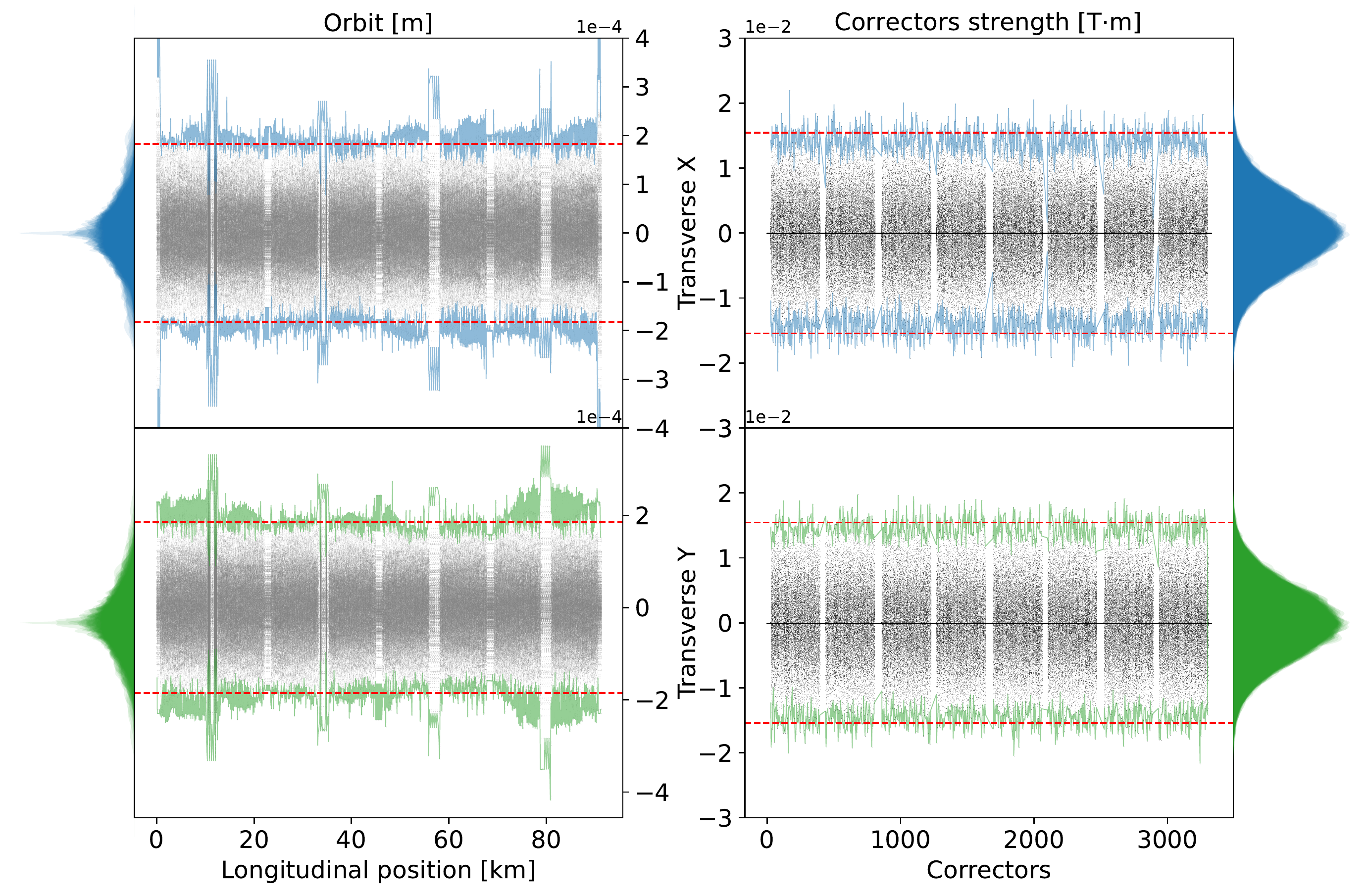}
    \caption{Superposition of the orbits the and correctors strengths of 99 seeds (dots) and the global envelope for all considered machine configurations (solid line) in the X and Y transverse planes. Density distributions (one for each seed) are also superposed. The red dashed line is 3 times the mean of the analytical RMS values.}
    \label{fig:orb_cor}
\end{figure}

The figure~\ref{fig:orb_cor} shows the orbit and correctors strengths  values and distributions with their respective analytical RMS values for the 99 successful seeds at the end of the correction procedure. The dashed red lines on the distributions represents the RMS calculated analytically. 

In both the orbit and the corrector strength analyses, we observe that our analytical predictions align well with the simulation data. Specifically, the estimation accurately captures all the data for the vertical plane, while the horizontal plane has a few outliers that exceed the analytical limit. These behaviors can be explained by the combined effect of different errors and the $\beta$-function, which renders the procedure less effective. Nonetheless, the orbit distribution corresponds to our expectations, with the amplitude in both planes being in the order of magnitude of the MQ offsets (the dominant errors). Moreover, the pattern of the succession of the arcs and insertions is apparent, as we only applied the errors on the arcs, and the residual orbit after correction in the insertions should be almost zero.
The RMS values for each successful seed are distributed around the dashed red line representing one analytical RMS estimate. The blue dots correspond to the RMS after turning on the sextupoles and correcting the orbit once, always using a Singular Value Decomposition (SVD) algorithm. Few outlier seeds exist that do not improve with the iteration of the corrector procedure.

Regarding the correctors' strength distribution (see figure~\ref{fig:orb_cor}), we observe that the strengths are contained in the 3 times mean analytical RMS limit. Therefore, the first specifications for the main magnets' misalignment of the High Energy Booster arcs cells are set to 150 $\mu$m, with a maximum corrector strength of about 20 mT$\cdot$m, as far as orbit correction is concerned. These values will be confirmed or potentially reduced following the full emittance tuning performance study.

\section{Collective effects and injection parameters} 
The high energy booster's smaller pipe radius of 25 mm, compared to the collider's 35 mm, and almost 100 km circumference, along with the required injection energy and bunch population for the desired emittance at different operating modes (shown in table \ref{tab:rf_budget}), make collective effects important. Two of these effects being investigated are the resistive wall and intrabeam scattering (IBS).

As for the resistive wall, because of the reduced pipe diameter, the  impedance and the wakefield contributions are expected to be higher than the main rings \cite{PhysRevAccelBeams.21.041001, refId0} in both longitudinal and transverse plane\footnote{these two effects scale as $r^{-1}$ in the longitudianl plane and as $r^{-3}$ in the transverse one.}.

\begin{figure}[!h]%
\centering
\includegraphics[width=0.49\columnwidth]{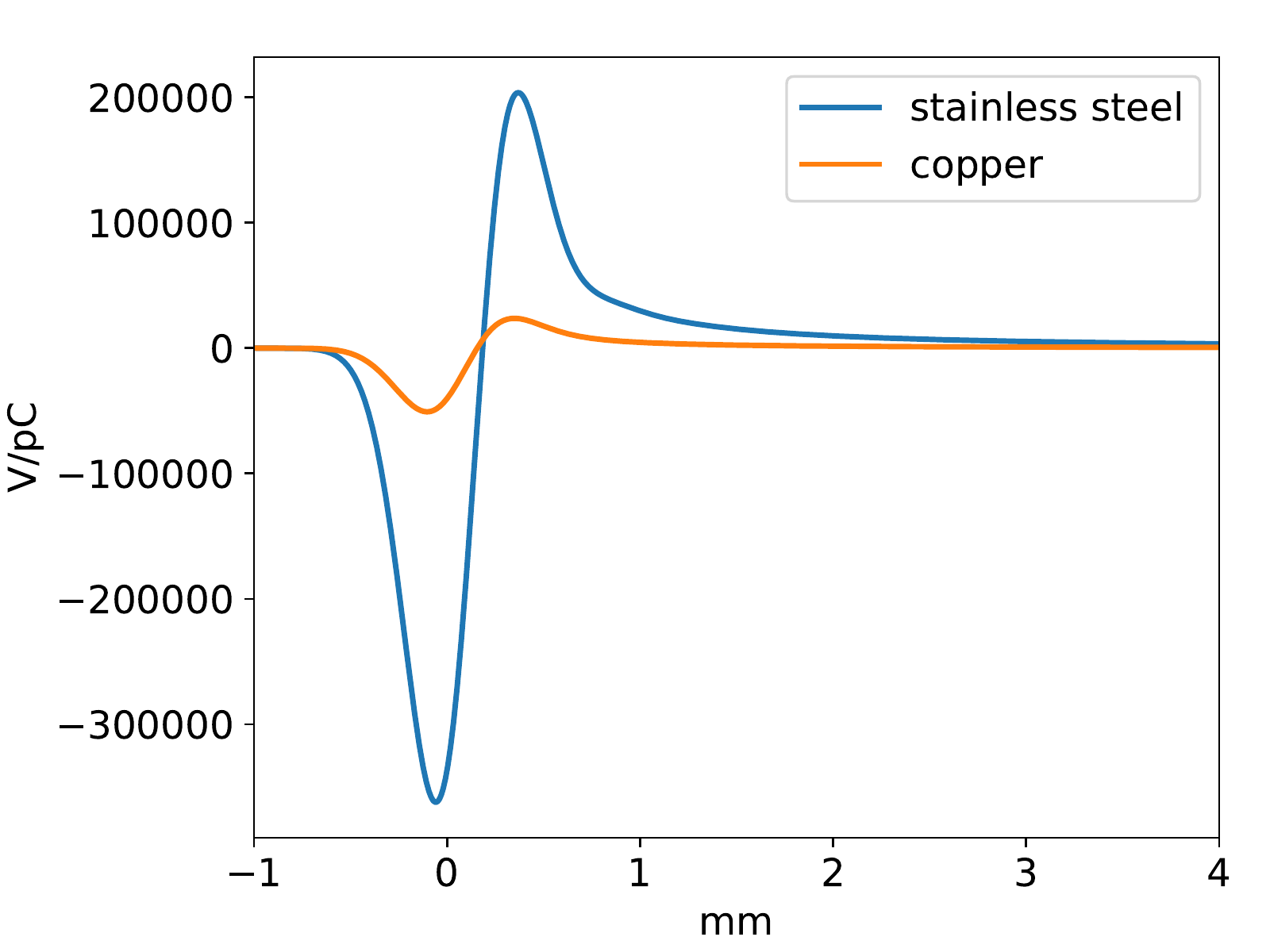}
\includegraphics[width=0.49\columnwidth]{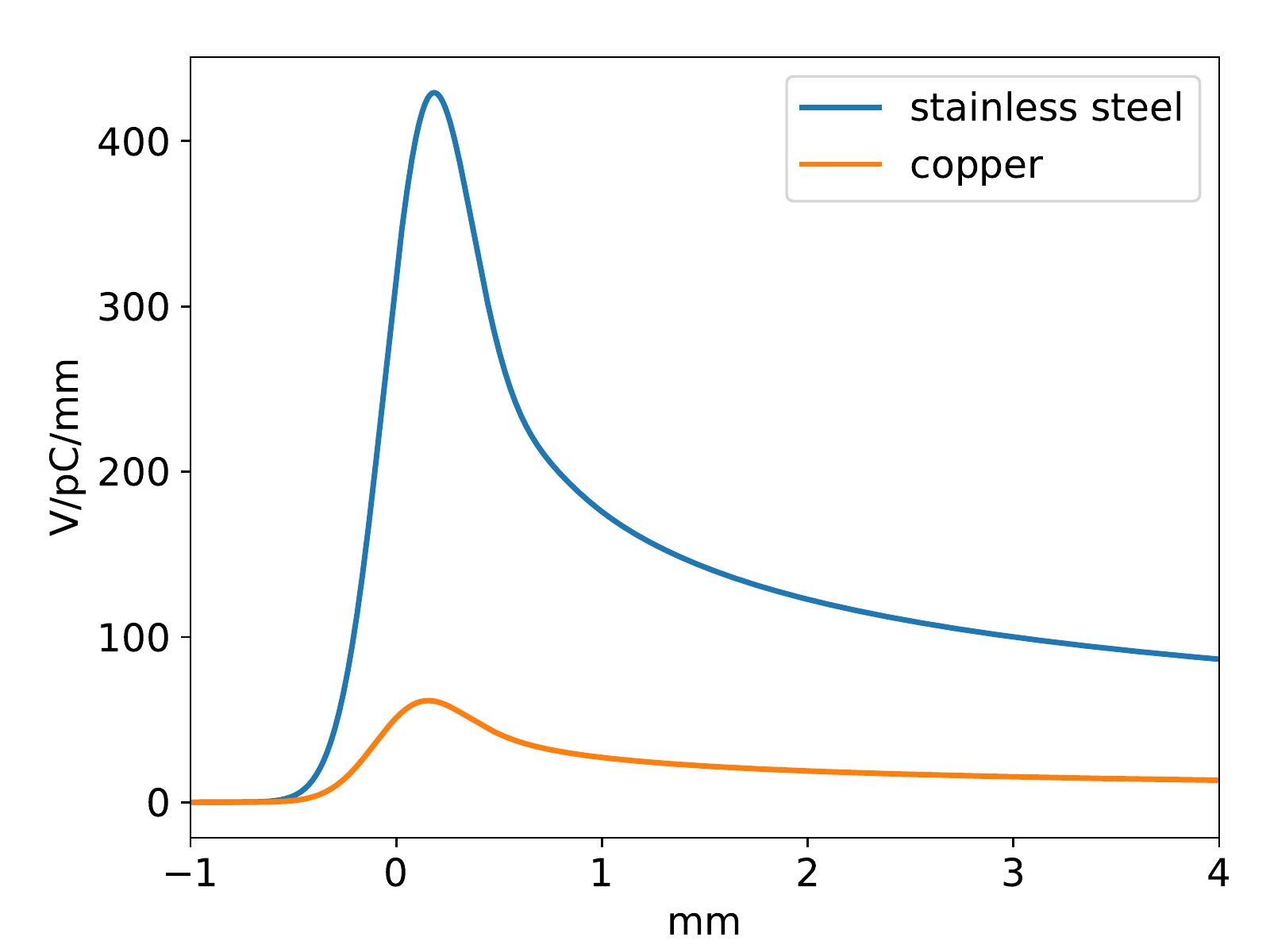}
    \caption{Illustration of longitudinal (left) and transverse dipolar (right) wake potential of a 0.2 mm Gaussian bunch for a copper and stainless steel vacuum chamber.}%
    \label{fig:wakes}%
\end{figure}

Another important point related to the resistive wall is the possible presence of eddy currents that could be generated by the ramp of the magnetic field during acceleration. In order to avoid them, an initial proposal of a stainless steel vacuum chamber was discussed. This material would have increased the resistive wall contribution by a too large amount. However, due to the low magnetic field in the booster \footnote{according to the CDR \cite{benedikt2018fcc} from 0.005 T at 20 GeV to a maximum peak of 0.046 T at 182.5 GeV, with a rate of field change below 0.03 T/s.}, the eddy currents are not expected to be a problem, giving, as consequence, that it is possible to have a stainless steel vacuum chamber with a copper coating of 1 mm, so that, from about 2 KHz on, the skin depth is such that the beam sees only the copper. However, one simple way to fabricate such a vacuum chamber would give, as a final result, a copper pipe with a stainless steel strip about 1 mm wide\footnote{Private communication.} seen by the beam. This strip would produce an azimuthal asymmetry in the impedance and this contribution, also taking into account the detuning (quadrupolar) terms, has to be studied in detail. In this section, we present only preliminary results for a circular beam pipe made of copper. The longitudinal and transverse wake potentials of a 0.2 mm Gaussian bunch used as a pseudo-Green function for copper and stainless steel are shown in figure~\ref{fig:wakes}.

With such contributions, the effects on beam dynamics have been studied with the PyHEADTAIL tracking code \cite{li2016code}. 
As initial conditions for the bunch distribution we have supposed a transverse emittance of $3\times 10^{-4}$ mm$\cdot$mrad in the horizontal plane and 1.4 $\mu$m$\cdot\mu$rad in the vertical plane, with an equilibrium relative energy spread of $3 \times 10^{-4}$ and a zero current bunch length of about 2.2 mm.

\begin{figure}[!h]%
\centering
\includegraphics[width=\columnwidth]{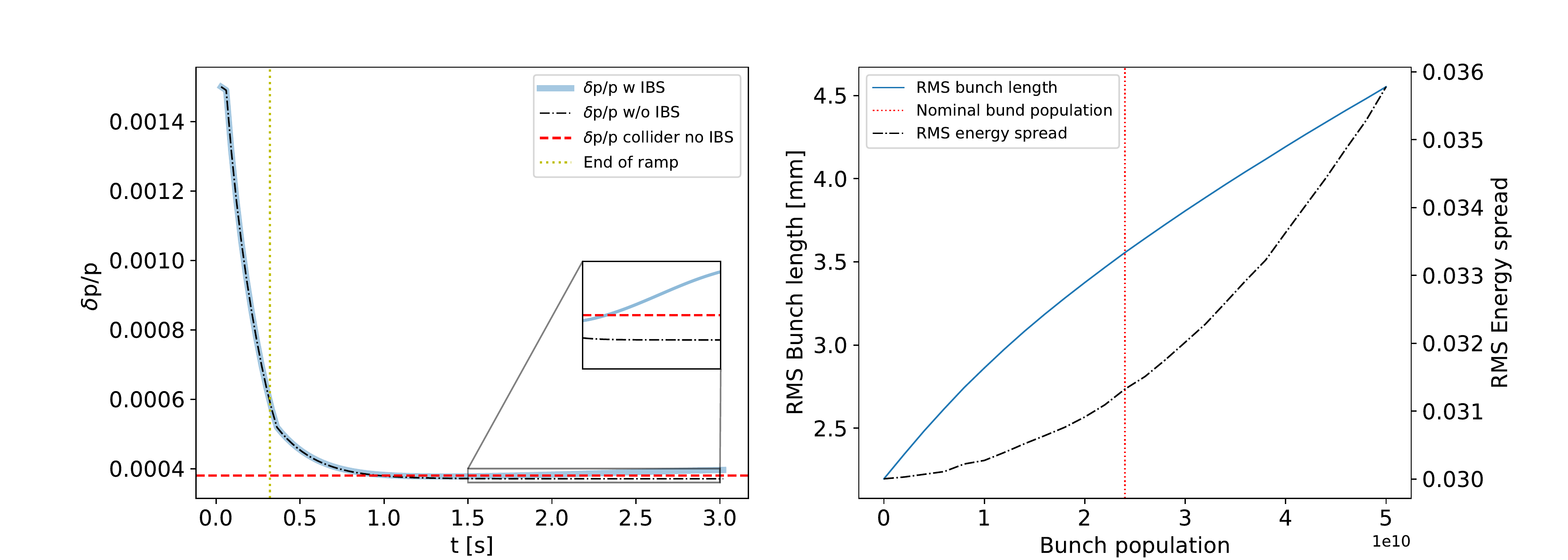}
    \caption{Left : Effect of the IBS on the evolution of the energy spread during a linear  ramp for an injected beam energy spread and bunch length respectively set to 0.15\% and 1 mm. Right : RMS bunch length and and energy spread versus bunch population.}
    \label{fig:sigma_vs_current}%
\end{figure}

The effect of the longitudinal wakefield is shown in figure~\ref{fig:sigma_vs_current}. We can see that, at the nominal bunch population of $2.4\times 10^{10}$ we have an effect of the potential well distortion, but also a slight microwave instability which increases the energy spread by a few percent with respect to its nominal value. 

\begin{figure}[!h]%
\centering
\includegraphics[width=0.7\columnwidth]{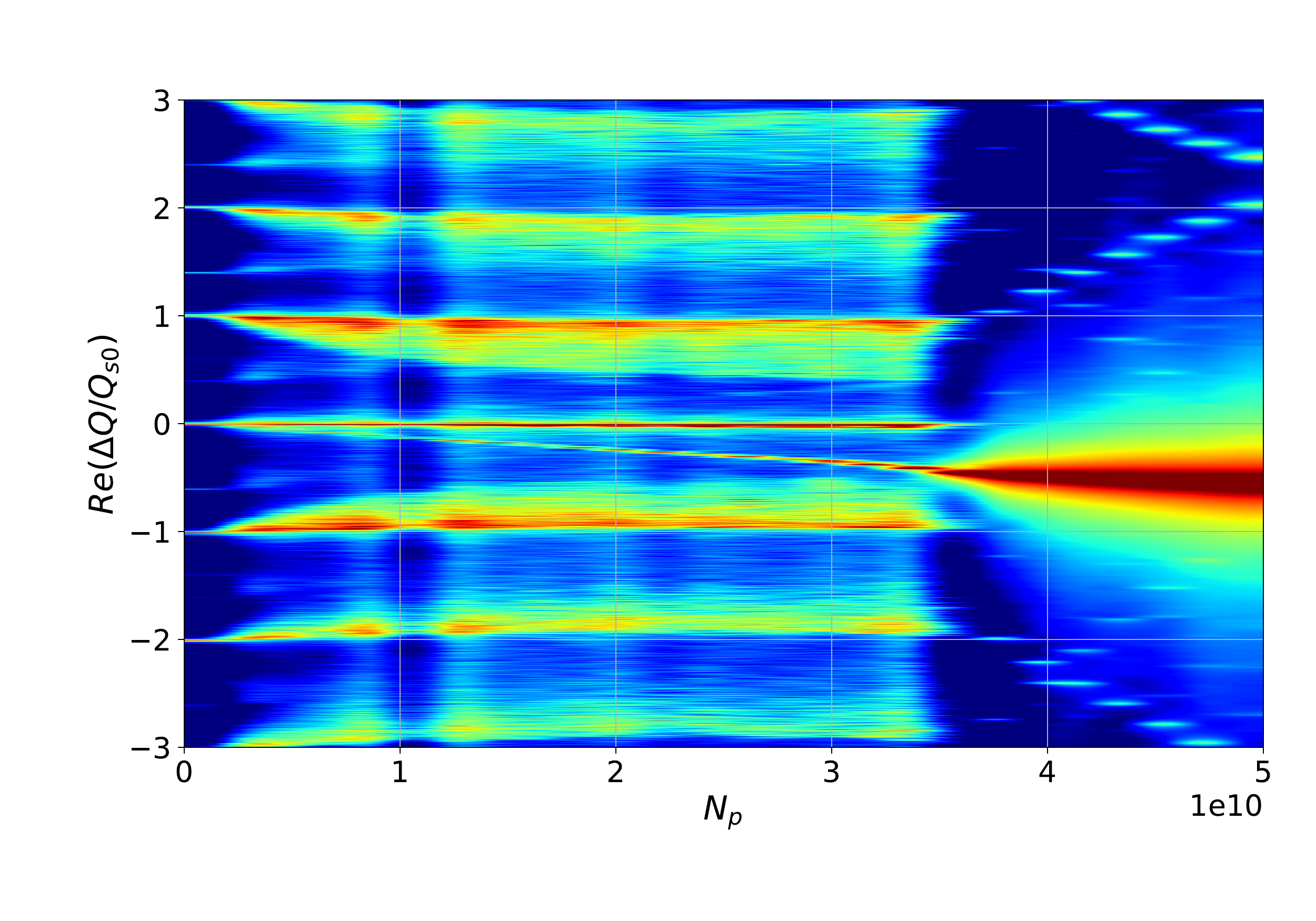}
    \caption{Real part of the tune shift of the ﬁrst azimuthal transverse coherent oscillation modes normalised by the synchrotron tune $Q_{s0}$ as a function of bunch population.}%
    \label{fig:tmci}%
\end{figure}

In the transverse plane, the wakefield can produce Transverse Mode Coupling Instability (TMCI), which can be more dangerous than microwave instability because it can cause beam loss. The instability occurs when two coherent oscillations modes couple together. As shown in figure~\ref{fig:tmci} \cite{PhysRevAccelBeams.23.071001}, the threshold is higher than the nominal current, but the safety margin is not so large.

It is important to note that the presented results provide only a rough evaluation of the collective effects on the beam dynamics of the booster. Self-consistent simulations that take into account the intrabeam scattering and wakefields, both influencing the bunch length, must be considered. Additionally, the possible difference in the resistive wall impedance due to the presence of the strip, which produces also an asymmetry with contributions to the detuning (quadrupolar) impedance, must be carefully evaluated. Finally, the machine impedance model has to be built. For the IBS, an updated analytical study made with MadX with respect to the one published in \cite{dalena2022status} is represented in figure \ref{fig:sigma_vs_current}. It shows an important contribution of the IBS after energy ramp to the energy spread making it higher than the collider threshold. This result is however to be confirmed with tracking simulation taking into account simultaneously the wakefield and the IBS.

\section{RF parameters and frequency choice}
As the HEB is central in the accelerator complex design, it provides inputs to the different working group studies. One of these inputs is the radiofrequency (RF) total voltage budget, which depends, among other things, on the bunch length needed at extraction for different energy modes and the RF frequency of the accelerating cavities. Recently, the base design of the RF frequency has changed from 400 MHz to 800 MHz, requiring a revision of the total voltage budget. Assuming no energy gain ($E_{gain}=0$), one can calculate the resulting RF voltage using the Eq~\eqref{eq.Vrf}:

\begin{equation}
    V_{RF} = \sqrt{
    \left(
    \frac{C^2\sigma_e^2 E_t \eta}{2\pi\nu_{RF}\sigma_{z}^2\beta^3}
    \right)^2 + \left(E_{gain}+U_{0}\right)^2}
\label{eq.Vrf}\end{equation}

In this equation, C represents the booster circumference, $\sigma_z$ is the bunch length, $E_t$ is the total energy, $\eta\simeq -\alpha_c$ is the slippage factor, $\beta$ is the normalized velocity, $\nu_{RF}$ is the RF cavities frequency, $U_0$ is the synchrotron energy loss per turn, and $\sigma_e$ is the energy spread.

At injection energy, the momentum acceptance is taken as a criterion for the calculation of the cavities RF voltage budget by solving Eq. \eqref{eq:dp}:
\begin{equation}
    \delta_p = \frac{2 Q_s(V_{RF}) c}{C \nu_{RF} \alpha_c}\sqrt{\tan\phi_s(V_{RF})*(1+\phi_s(V_{RF}) - \pi/2)}
\label{eq:dp}\end{equation}
with $Q_s(V_{RF})$ the synchronous tune and $\phi_s(V_{RF})$ the synchronous phase.

 As can be seen in table \ref{tab:rf_budget}, the frequency change from 400 MHz to 800 MHz allows to reduce the total cavities RF budget required at extraction. However, taking the momentum acceptance as a requirement at injection almost doubles it when the RF frequency is doubled.
 
\begin{table}
\caption{Total RF voltage budget for the different energy modes of the FCC-ee HEB for two RF cavities frequencies.}\label{tab:rf_budget}
\begin{minipage}[b]{\textwidth}\footnotesize
\begin{tabular*}{\textwidth}{@{\extracolsep{\fill} }l*{5}{c}r}
\toprule
Modes  & Injection $\ang{60}$\footnote{$\ang{60}$ phase advance lattice of the Z and W modes}/$\ang{90}$\footnote{$\ang{90}$ phase advance lattice of the H and $t\bar{t}$ modes} &Z & W & H & $t\bar{t}$ & \scriptsize{[Units]}\\
\midrule
Energy	& 20 & 45.6  & 80 & 120 & 182.5 & \scriptsize{[GeV]}\\
\midrule
$\sigma_z$\footnote{Bunch length} & 4 & 4.38 & 3.55 & 3.34 & 1.94 & \scriptsize{[mm]}\\
$\delta_p$\footnote{Momentum acceptance} & 3 & 4.38 & 3.55 & 3.34 & 1.94 & \scriptsize{[mm]}\\
$\alpha_c$\footnote{Momentum compaction factor} & 14.9 / 7.34 & \multicolumn{2}{c}{14.9} & \multicolumn{2}{c}{7.34} & \scriptsize{[$10^{-6}$]}\\
\midrule
$V_{RF,400}$\footnote{RF voltage for $\nu_{RF}=400~MHz$} & 53.6 / 27.6 & 124.6 & 1023.2 & 2185.6 & 14205.4 & \scriptsize{[MV]}\\
$V_{RF,800}$\footnote{RF voltage for $\nu_{RF}=800~MHz$} & 104.8 / 52.8 & 83.9 & 623.6 & 2038.3 & 11554.9 & \scriptsize{[MV]}\\
\bottomrule
\end{tabular*}
\end{minipage}
\end{table}

\section{Alternative Optics}
In the baseline, the arcs are made of FODO cells with a phase advance of $\ang{60}$/$\ang{90}$ respectively for the modes $Z$/$W$ and $H$/$t\bar{t}$. The main reason is a larger momentum compaction at injection at the $Z$/$W$ modes to manage stronger collective effects due to a larger current. However, to enlarge the dynamic aperture, the sextupole scheme is based on a non-interleaved scheme with a phase advance of $\ang{180}$ between two sextupoles of a pair.
Maintaining the possibility to have $\ang{60}$/$\ang{90}$ implies to have a different cabling for the different operation modes. Moreover, the number of sextupoles is roughly doubled.

\begin{figure}[!h]%
    \centering
    \includegraphics[width=0.8\columnwidth]{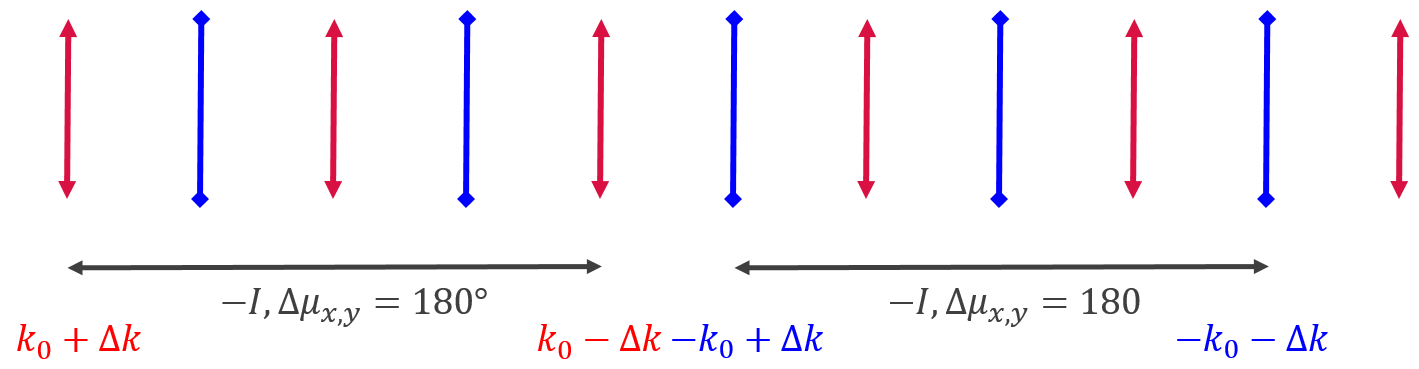}
    \caption{Sheme to tune the momentum compaction. The focusing/defocusing quadrupoles are respectively in red/blue.}
    \label{fig:alpha_tune_scheme}%
\end{figure}

We propose an alternative scheme which enables to tune the momentum compaction by keeping the same non-interleaved scheme for all operation modes.
The principle is to create a dispersion and betatron wave at one quadrupole near a sextupole.
We assume for instance that the integrated quadrupole strength is modified by $\Delta k$; the quadrupole near the other paired sectupole is modified by $-\Delta k$ (see figure~\ref{fig:alpha_tune_scheme}).
The phase advance between the center of the 2 quadrupoles is $\ang{180}$ in both planes.
By this way, the betatron wave is cancelled in the second quadrupole contrary to the dispersion wave (because the frequency of the betatron wave is twice the one of the dispersion wave).
The other advantage is also not to change the tune of the cell.
If we do the thin lens approximation then to get a relative change of $x$ on the momentum compaction, we should have $\Delta k \approx \frac{\sqrt{x}}{2\sqrt{3}}$

The advantages of this scheme are to simplify the cabling and distribution of the sextupoles in the arcs; to enable to tune the momentum compaction during the ramp and thus to get a smaller equilibrium emittance for the $Z$/$W$ modes; to be compatible with any arc pattern if the phase advance between the sextupoles is still \ang{180}. The drawbacks are an additional power supply for the quadrupoles (but that should be less expensive than increasing the number of sextupoles if we have to keep non-interleaved scheme for \ang{60}/\ang{90} cells), a larger equilibrium emittance in comparison with a FODO cell giving the same momentum compaction, a possible reduction of the dynamic aperture and momentum acceptance.

\section{Conclusion}
This paper reports on the status of several ongoing parallel developments for the FCC-ee High Energy Booster. One of the ongoing studies focuses on particle stability in longitudinal phase space, particularly with regard to the control of synchro-betatron excitations. Other investigations are aimed at evaluating the performance of the current HEB optics design, in particular with respect to stability against lattice elements imperfections, main sextupoles settings, and the interplay of collective effects with a realistic lattice model. By integrating Wakefield and IBS, together with synchrotron radiation, we aim to gain important insights into the injection parameter constraints. These simulations will also allow us to compare the effects of different energy ramping strategies.

All these studies will provide input for alternative optics design and RF parameter choices for the different operating modes, while maintaining compatibility with both pre-injectors and collider requirements. However, it is important to note that these studies are highly dependent on various inputs such as accelerating cavities RF frequency, bunch population, vacuum pipe geometry, material and coating, and collider length, among other parameters. Therefore, they must be conducted in parallel and in dynamic interaction with other ongoing studies such as pre-injectors, collider optics, radio-frequency, or arcs optics patterns to ensure that all the necessary inputs are taken into consideration, and that the most optimal solutions are achieved.

\section*{Acknowledgements}
This work was supported by the European Union’s Horizon 2020 Research and Innovation programme under grant no.~951754 (FCCIS).

\bibliographystyle{JHEP}
\bibliography{biblio.bib}

\begin{thebibliography}{8}
\providecommand{\natexlab}[1]{#1}
\providecommand{\url}[1]{\texttt{#1}}
\expandafter\ifx\csname urlstyle\endcsname\relax
  \providecommand{\doi}[1]{doi: #1}\else
  \providecommand{\doi}{doi: \begingroup \urlstyle{rm}\Url}\fi

\bibitem[Benedikt et~al.(2018)Benedikt, Mertens, Zimmermann, Cerutti, Otto,
  Poole, Brunner, Gutleber, Milanese, Blondel, et~al.]{benedikt2018fcc}
Michael Benedikt, Volker Mertens, Frank Zimmermann, Francesco Cerutti, Thomas
  Otto, John Poole, Olivier Brunner, Johannes Gutleber, Attilio Milanese, Alain
  Blondel, et~al.
\newblock Fcc-ee: The lepton collider: Future circular collider conceptual
  design report volume 2.
\newblock \emph{Eur. Phys. J. Spec. Top.}, 228\penalty0
  (CERN-ACC-2018-0057):\penalty0 261--623, 2018.

\bibitem[Dalena et~al.(2022)Dalena, Chance, Antoniou, Etisken, Mashal,
  Raubenheimer, Zampetakis, and Zimmermann]{dalena2022status}
B.~Dalena, A.~Chance, F.~Antoniou, O.~Etisken, A.~Mashal, T.~Raubenheimer,
  M.~Zampetakis, and F.~Zimmermann.
\newblock {Status of the High Energy Booster of the lepton option of the future
  circular collider}.
\newblock \emph{PoS}, ICHEP2022:\penalty0 042, 11 2022.
\newblock \doi{10.22323/1.414.0042}.

\bibitem[Wiedemann(2015)]{wiedemann}
Helmut Wiedemann.
\newblock \emph{Particle accelerator physics}.
\newblock Springer Nature, 2015.

\bibitem[Dalena et~al.(2023)Dalena, Da~Silva, Chance, and Ghribi]{ipac_orbcor}
Barbara Dalena, Tatiana Da~Silva, Antoine Chance, and Adnan Ghribi.
\newblock Definition of tolerances and corrector strengths for the orbit
  control of the high-energy booster ring of the future electro-positron
  collider.
\newblock In \emph{14th International Particle Accelerator Conference
  (IPAC2023)}, 2023.

\bibitem[Migliorati et~al.(2018)Migliorati, Belli, and
  Zobov]{PhysRevAccelBeams.21.041001}
M.~Migliorati, E.~Belli, and M.~Zobov.
\newblock Impact of the resistive wall impedance on beam dynamics in the future
  circular ${e}^{+}{e}^{\ensuremath{-}}$ collider.
\newblock \emph{Phys. Rev. Accel. Beams}, 21:\penalty0 041001, Apr 2018.
\newblock \doi{10.1103/PhysRevAccelBeams.21.041001}.
\newblock URL
  \url{https://link.aps.org/doi/10.1103/PhysRevAccelBeams.21.041001}.

\bibitem[{Migliorati, M.} et~al.(2021){Migliorati, M.}, {Carideo, E.}, {De
  Arcangelis, D.}, {Zhang, Y.}, and {Zobov, M.}]{refId0}
{Migliorati, M.}, {Carideo, E.}, {De Arcangelis, D.}, {Zhang, Y.}, and {Zobov,
  M.}
\newblock An interplay between beam-beam and beam coupling impedance effects in
  the future circular e+e- collider.
\newblock \emph{Eur. Phys. J. Plus}, 136\penalty0 (11):\penalty0 1190, 2021.
\newblock \doi{10.1140/epjp/s13360-021-02185-2}.
\newblock URL \url{https://doi.org/10.1140/epjp/s13360-021-02185-2}.

\bibitem[Li et~al.(2016)Li, Bartosik, Iadarola, Oeftiger, Passarelli, Romano,
  Rumolo, Schenk, and Hegglin]{li2016code}
Kevin Li, H~Bartosik, G~Iadarola, A~Oeftiger, Andrea Passarelli, A~Romano,
  G~Rumolo, M~Schenk, and S~Hegglin.
\newblock Code development for collective effects.
\newblock In \emph{ICFA Advanced Beam Dynamics Workshop on High-Intensity and
  High-Brightness Hadron Beams (HB2016)}, 2016.

\bibitem[M\'etral and Migliorati(2020)]{PhysRevAccelBeams.23.071001}
E.~M\'etral and M.~Migliorati.
\newblock Longitudinal and transverse mode coupling instability: Vlasov solvers
  and tracking codes.
\newblock \emph{Phys. Rev. Accel. Beams}, 23:\penalty0 071001, Jul 2020.
\newblock \doi{10.1103/PhysRevAccelBeams.23.071001}.
\newblock URL
  \url{https://link.aps.org/doi/10.1103/PhysRevAccelBeams.23.071001}.

\end{thebibliography}

\end{document}